\documentclass[prd,twocolumn,reprint,preprintnumbers,nofootinbib,superscriptaddress,longbibliography]{revtex4-1}
\usepackage{amsmath}
\usepackage{amssymb} 
\usepackage{enumerate}
\usepackage{multirow}
\usepackage{braket}
\usepackage[toc,page]{appendix}
\usepackage{url}
\usepackage{graphicx}
\usepackage{hyperref}
\hypersetup{colorlinks=true}
\usepackage{latexsym}
\usepackage{wasysym}

\usepackage{mathrsfs,amssymb}  
\usepackage{cancel}
\usepackage[normalem]{ulem}
\usepackage{cleveref}
 \usepackage{tikz}
\usetikzlibrary{decorations.markings}
\usetikzlibrary{positioning}
\usetikzlibrary{shapes}
\usetikzlibrary{calc}
\usepackage{textcomp}
\usepackage{lipsum}
\usepackage{newtxtext, newtxmath}
\usepackage[export]{adjustbox}
\usepackage{tabularray}
\UseTblrLibrary{booktabs} 

\usepackage{hhline}% http://ctan.org/pkg/hhline
\usepackage{soul}
\newcommand{\GeV}{{\,\rm GeV}}
\newcommand{\TeV}{{\,\rm TeV}}

\def\cleardoublepage{%
  \clearpage
  \if@twoside
    \ifodd\c@page
    \else
    \hbox{}\newpage
      \if@twocolumn\hbox{}\newpage\fi
    \fi
  \fi}

\begin{document} 
\preprint{KIAS-P23059}

\title{  
Exploring the Synergy of Kinematics and Dynamics  for Collider Physics 
 }

\author{Kayoung Ban}
\email{ban94gy@yonsei.ac.kr}
\affiliation{Department of Physics and IPAP, Yonsei University, Seoul 03722, Republic of Korea}
 
\author{Kyoungchul Kong}
\email{kckong@ku.edu}
\affiliation{Department of Physics and Astronomy, University of Kansas, Lawrence, KS 66045, USA}
 
\author{Myeonghun Park}
\email{parc.seoultech@seoultech.ac.kr}
\affiliation{School of Natural Sciences, Seoultech, Seoul 01811, Republic of Korea}

\author{Seong Chan Park}
\email{sc.park@yonsei.ac.kr}
\affiliation{Department of Physics and IPAP, Yonsei University, Seoul 03722, Republic of Korea}
\affiliation{School of Physics, Korea Institute for Advanced Study, Seoul 02455, Korea}
\date{November 27, 2023}
%%%%%%%%%%%%%%%%%%%%%%%%%%%%%%%%%%%%%%%%%%%%%%%%%%%%%%%%%%%%%%%%%%%%%%%%%%%%%%%%%%%%%%
\begin{abstract}
In collider experiments, an event is characterized by two distinct yet mutually complementary features: the `global features' and the `local features'. 
Kinematic information such as the event topology of a hard process, masses, and spins of particles comprises global features spanning the entire phase space. This global feature can be inferred from reconstructed objects.
 In contrast, representations of particles in gauge groups, such as Quantum Chromodynamics (QCD), offer localized features revealing the dynamics of an underlying theory. These local features, particularly observed in the patterns of radiation as raw data in various detector components, complement the global kinematic features. 
In this letter, we propose a simple but effective neural network architecture that seamlessly integrates information from both kinematics and QCD to enhance the signal sensitivity at colliders.
\end{abstract}
\maketitle
\flushbottom
%%%%%%%%%%%%%%%%%%%%%%%%%%%%%%%%%%%%%%%%%%%%%%%%%%%%%%%%%%%%%%%%%%%%%%%%%%%%%%%%%%%%%%
\section{Introduction}\label{sec:intro}
The Large Hadron Collider (LHC) has ushered in an era of data proliferation, characterized by unprecedented complexity and volume, which accelerated new physics searches beyond the Standard Model.
Amidst this abundance of data, Deep Neural Networks (DNNs) have risen as formidable and essential tools \cite{Albertsson:2018maf, Schwartz:2021ftp, Feickert:2021ajf, Radovic:2018dip, Shanahan:2022ifi,  Karagiorgi:2021ngt}, adept at unraveling the intricate correlations veiled within the vast, multi-dimensional data emanating from particle collisions \cite{Kim:2021pcz, Ban:2022hfk, Dong:2022trn, Franceschini:2022vck}.

\begin{figure}[t!] 
\includegraphics[width=0.47\textwidth]{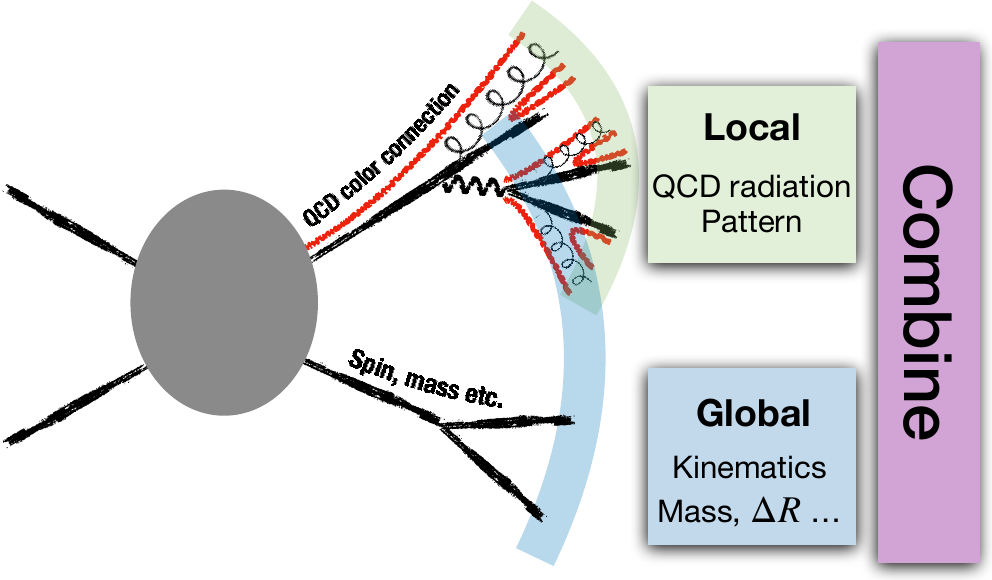}
\caption{The schematic overview of the types of information at a collider. A meticulous combining methodology is crucial, preserving and leveraging independent features to understand an underlying physics.}
\label{fig:main} 
\end{figure}

In collider phenomenology, kinematic features are characterized by global stance as depicted through various observables such as invariant mass, angular distance, and transverse momentum of final state particles encapsulating the event characteristics.
They capture the comprehensive dynamics of particle collisions, offering insights, and intrinsic attributes reflective of the high-energy phenomena involved \cite{Han:2005mu, Barr:2011xt, Franceschini:2022vck}.
Concurrently, local features, in particular, come from  activities of QCD interactions and capture the aspects of SU(3) gauge interactions. 
These local attributes predominantly pertain to the QCD color charges of quarks and gluons~\cite{Matthew:2010sw, deOliveira:2015xxd, Gallicchio:2010dq, Kagan:2020yrm, Komiske:2016rsd, Han:2023djl}. 
One often utilizes combinations of the global (kinematic) and the local (QCD specific) features in analyzing a collision event at the LHC, which is illustrated in FIG.~\ref{fig:main}.

To effectively combine these data types, a multi-modal neural network has been utilized, which generally employs two main approaches.
The first approach is focused on constancy, identifying correlations within the data to understand how different data types relate and influence each other, thereby enhancing the network's predictive accuracy through consistent patterns across modalities \cite{Zhang_Zhang_Pan_2022, Yuan_Cai_Wang_Li_2023,10230895,fei2022towards}. 
The second approach centers on complementarity, acting on independent information from each data type to provide a more complete picture \cite{KIM2022290, LIU2023679, KIM2022102501, 9844446}. 
This method enriches the overall analysis by integrating these distinct yet complementary data points, leading to a more nuanced understanding.
In the research area of high-energy physics, we adopt a complementary approach rather than consistency to maximize the use of information.
The optimization of combinations through complementary is challenging, as it often sensitively depends on the specific data set and underlying physics, an aspect not significantly emphasized in previous studies\,\cite{Lin:2018cin, Kim:2019wns, Esmail:2023axd}.

\begin{figure*}[t!]
    \centering
    \includegraphics[width=0.8\textwidth]{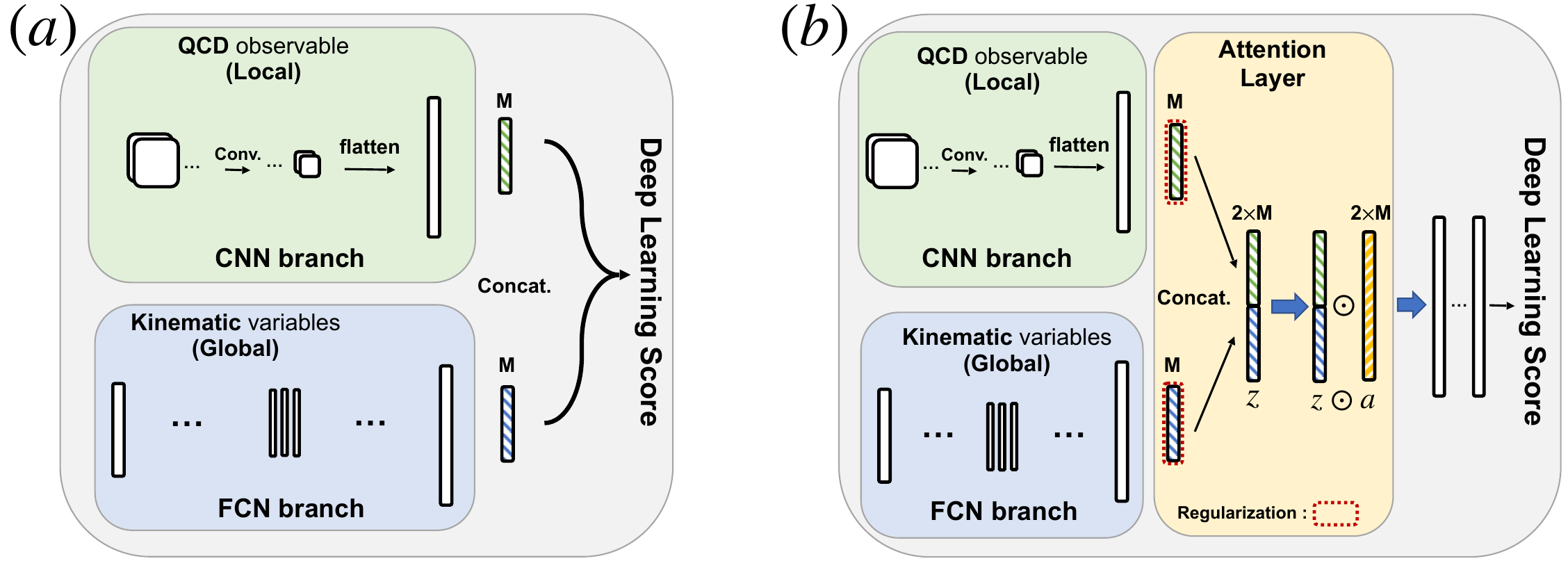}
    \caption{The schematic plots for neural network structures: (a) conventionally used one in previous studies only with concatenation and (b) our proposed one with a regularized attention mechanism.}
    \label{fig:model}
\end{figure*}

In this paper, we propose a novel approach integrating kinematic variables and QCD features in a multi-modal neural network in a complementary way. 
The focus is to investigate and mitigate the unintended shadowing effects that can occur in multi-modal deep neural networks when integrating local and global features. 
Our architecture is structured to maintain the fidelity of both kinematic variables and QCD features, ensuring that each data type is complemented and without shadowing.
To navigate the intricate interplay between the different types of particle data—encompassing global kinematic variables to local QCD features—a model capable of handling such complexity with precision is necessary.    

%%%%%%%%%%%%%%%%%%%%%%%%%%%%%%%%%%%%%%%%%%%%%%%%%%%%%%%%%%%%%%%%%%%%%%%%%%%%%%%%%%%%%%
\section{Optimizing Combination} \label{sec:network}
%\section{Network structure} \label{sec:network}
To extract information from kinematic variables, we introduce a Fully Connected Network (FCN) that takes multi-dimensional vector-type data as input.
Since the energy deposits of these jet particles can be represented as calorimeter images in the ($\eta$, $\phi$) plane, we utilize a simple but effective Convolutional Neural Network (CNN) that takes images as input.
We employ the following three models as a comparative set for our study:
\begin{itemize}
    \item[(a)] FCN with kinematic variables
    \item[(b)] FCN + CNN with QCD observables (1) without and (2) with Attention layer
    \item[(c)] FCN + CNN with QCD observables after Riemannian preprocessing \footnote{We will introduce the Riemannian preprocessing applied in our study in the following section \ref{sec:example}.} (1) without and (2) with Attention layer
\end{itemize}
We provide schematic diagrams of models (1) without and (2) with the attention layer in FIG. \ref{fig:model}.
The FCN model (a) solely considers kinematic variables. 
It comprises four fully connected layers: each 
of which utilizes ReLU activation and is interspersed with batch normalization. The model is completed with an output layer with a sigmoid activation function.
The models (b) and (c), as multi-modal designs, will be detailed in the following.

%% CNN structure %%%%%%%%%%%%%%%%%%%%%%%%%%%%%%%%%%%%%%%%%%%
\textit{CNN branch (Local features)}:
The CNN branch accepts 2D arrays of shape 32×32×2 as input. 
Two channels consist of images of charged particles and neutral particles, respectively. This is in accordance with the previous study \cite{Kim:2019wns}.

%% FCN structure %%%%%%%%%%%%%%%%%%%%%%%%%%%%%%%%%%%%%%%%%%%
\textit{FCN branch (Global features):} 
The FCN branch takes a 1D array of $N$ kinematic variables as inputs.

%% merged %%%%%%%%%%%%%%%%%%%%%%%%%%%%%%%%%%%%%%%%%%%
\textit{Merging}:
In the previous studies  \cite{Kim:2019wns, Huang:2022rne, Flacke:2023eil, Hammad:2022lzo, Esmail:2023axd}, information from kinematics and QCD is simply concatenated in classification models. 
However, simply concatenating the outputs from the FCN and CNN branches often results in an imbalance. 
This concatenation approach often results in a dominance of kinematic features derived from the FCN, which can inadvertently overshadow the valuable contributions of local features, such as QCD color structures. 
Also, multi-modal networks generally do not apply activation in the last layer of each branch to prevent loss of information.
However, this can cause the extracted layer to be an imbalanced data scale.
To resolve these issues, we introduce additional network structures.

%% Attention layer %%%%%%%%%%%%%%%%%%%%%%%%%%%%%%%%%%%%%%%%%%%
\textit{Attention layer}:
To mitigate this imbalance due to the non-activated layer, we employ regularization and an attention mechanism into a multi-modal framework.
Specifically, an L2 regularization term is added, governed by a hyperparameter $l_2$, to the loss function in the output layer of each individual model branch. 
This constraint is essential for preventing any single branch from dominating the multi-modal model, thereby ensuring that both local and global features 
contribute optimally to the final prediction.
This penalizes large weights, thereby encouraging the multi-modal model to merge as balanced with the two different types of model.
The L2 regularization term has the form of a lagrangian multiplier as
\begin{equation}
\text{Regularization term} = l_2 \times \sum_{k=1}^{M} W_k^2\, .
\end{equation}
$W_k$ denotes the weights of the output of FCN or CNN branch, and $l_2$ is a hyperparameter for a regularization.

To resolve the overshadowing problem, we define the attention score as $a$ with the softmax function  \cite{luong2015effective,bahdanau2014neural, 8237712, LIU2023679} 
\begin{equation}
    a(z) = \frac{e^{f_{\text{attn}}(z)}}{\sum_{j=1}^{2M} e^{f_{\text{attn}}(z)_j}}
    \label{eq:att}
\end{equation}
where the $z = [z_{1:M} : z_{M+1:2M}]$ is the concatenated layer from CNN and FCN respectively, and $f_{\text{attn}}(x)$ is a trainable linear transformation, $W_{\text{attn}}x+b_{\text{attn}}$.
Attention values are calculated as the element-wise product ($z \odot a$) between the attention score and the concatenated layer.
With above additions, our neural network can learn how to choose important information and it becomes more transparent by showing its focus on each branch in classifying a signal and a background.
The attention values are then connected with a fully connected layer for classification. 

%%%%%%%%%%%%%%%%%%%%%%%%%%%%%%%%%%%%%%%%%%%%%%%%%%%%%%%%%%%%%%%%%%%%%%%%%%%%%%%%%%%%%%
\section{Example: $HH$ vs $t\bar{t}$}
With the current interests of the LHC is being focused on the ``precision Higgs", we consider a process $pp\to HH$ at the LHC~\cite{Papaefstathiou:2012qe, Kim:2018cxf, Kim:2019wns, Huang:2017jws, Dolan:2012rv, Nakamura:2017irk, CMS:2017ums, Sirunyan_2018, ATLAS:2018fpd, Aad_2020, ATLAS:2022xzm}. This double Higgs production channel offers a unique probe into the Higgs boson self-coupling which is a key parameter in the Higgs potential.  
Accurate measurements of the Higgs self-coupling provide insights on electroweak symmetry breaking, stability of the electroweak vacuum, (critical) Higgs inflation and the potential for new physics beyond the Standard Model (SM)~\cite{Degrassi:2012ry, DeSimone:2008ei, Hamada:2014wna, Hamada:2014iga}.

Out of various decay modes in the double Higgs channel, we consider $\left(b\bar b \ell \bar \ell \nu \bar\nu\right)$ mode from $H\to b\bar b$ and $H\to W W^* \to \ell\bar\nu \bar \ell \nu$. 
The dominant background is $t\bar{t}$ as an irreducible one. This example serves as an exemplary case study for two reasons: firstly, the kinematic characteristics of the signal and backgrounds exhibit substantial differences; and secondly, the QCD radiation patterns from the $b\bar b$-system in both the signal and backgrounds offer distinctive features attributed to the disparate $SU(3)$ representations of the $H$ and the top quark. 

%%%%%%%%%%%%%%%%%%%%%%%%%%%%%%%%%%%%%%%%%%
\subsection{Event Selection}
To generate Monte Carlo samples, we use the standard chain of \textsc{MadGraph5\_aMC@NLO}\,\cite{Alwall:2014hca}, PYTHIA 8\,\cite{Sjostrand:2014zea}, and  Delphes\,\cite{deFavereau:2013fsa} for the $14\TeV$ LHC. Jet reconstruction is performed based on the anti-$k_T$ algorithm \cite{Cacciari:2008gp} with a radius parameter $ \Delta R = 0.4 $ and a transverse momentum threshold $ p_{Tj} > 20\GeV$. The $ b $-tagging efficiency is set at $ \epsilon_{b\rightarrow b} = 0.7 $, with a misidentification rate of $ \epsilon_{c\rightarrow b} = 0.2 $ and $ \epsilon_{j\rightarrow b} = 0.01 $. The baseline cuts are applied as follows. The transverse momentum $ p_T $ of the two leading $ b $-tagged jets is set to be greater than $30\GeV$, and two isolated leptons of opposite charge are required to have $ p_T $ greater than $20\GeV$. The missing transverse momentum should exceed $20\GeV$. The separation in the $\eta-\phi$ space for $ b\bar{b} $ pairs, represented by $ \Delta R_{b\bar{b}} $, is restricted to be less than 1.8, while for lepton pairs, $ \Delta R_{\ell\ell} $ should be under 1.3. 
The invariant mass of the lepton pairs $ m_{\ell\ell} $ is constrained to be less than $65\GeV$ and the invariant mass of $b$-tagged jets lies between 
$95 \GeV \le m_{b\bar{b}} \le 140\GeV$. 
Lastly, the pseudorapidity $\eta_{j}$ of the jets is capped at 2.5.

\subsection{Kinematic information}
We utilize conventional $N=10$ kinematic variables\,\cite{CMS-PAS-FTR-16-002,Kim:2019wns} 
\begin{itemize}
    \item The magnitude of the missing transverse momentum  
    \item Transverse momentum of each lepton with $p_T$ ordering, 
    \item The angular distance in the ($\eta,~\phi$) plane between $\ell^-$ and $\ell^+$, between $b$ and $\bar{b}$  
    \item Invariant mass of the $(\ell^-\ell^+)$ and $(b\bar{b})$ respectively,
    \item Transverse momentum of $(\ell^-\ell^+)$ and $(b\bar{b})$ respectively,
    \item The azimuthal angles between the $(\ell^-\ell^+)$ and $(b\bar{b})$.
\end{itemize}

%%%%%%%%%%%%%%%%%%%%%%%%%%%%%%%%%%%%%%%%%%
\subsection{QCD information and Decorrelation}\label{sec:example}
 
  There is a distinct difference in the radiation patterns emanating from jets produced by color singlet and color octet particles. QCD shower radiations from a color singlet particle tend to align more closely with the direction of the other jet due to color connection and color dipole effects, resulting in soft radiation that fills the space between the two quark jets. In contrast, radiations from color octet particles tend to spread out more broadly as they are predominantly directed toward the beam axis \cite{Matthew:2010sw, Gallicchio:2010dq, Han:2023djl}.
 
To enhance the complementary capabilities of multi-modal deep learning, we use a Riemannian mapping which is designed to remove the characteristics of the kinematic feature $\Delta R_{b\bar{b}}$ embedded in the jet image\,\cite{Hammad:2022msl}.
With this geometric decorrelation method, one can focus on the color connectivity without being distracted by the kinematics.
We demonstrate the results of decorrelation with global features using Principal Component Analysis (PCA) and linear regression methods in Appendix \ref{sec:appendix1}.
Finally, we take a conventional procedure in dealing with jet images  \cite{deOliveira:2015xxd, Kim:2019wns, Huang:2022rne} as in Appendix\,\ref{sec:appendixjet}.

\begin{figure}[t!]
    \centering
    \includegraphics[width=0.48\textwidth]{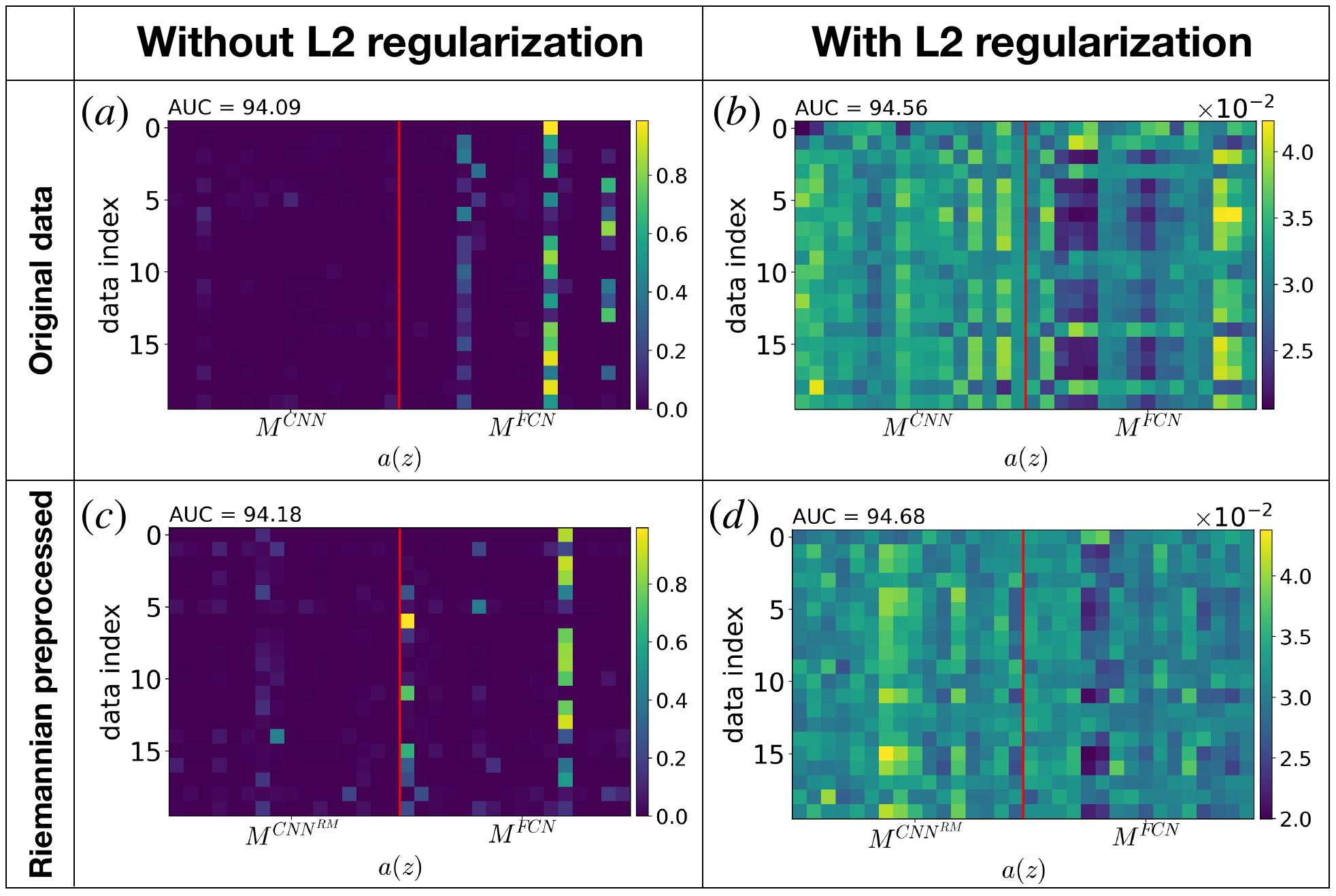} 
    \caption{\label{fig:L2diff}The comparison of attention scores $a(z)$ in $x$-axis for a test dataset comprising 20 samples with a data index in $y$-axis are evaluated post-training under two distinct neural network models using two different datasets. }
\end{figure}
%%%%%%%%%%%%%%%%%%%%%%%%%%%%%%%%%%%%%%%%%%%%%%%%%%%%%%%%%%%%%%%%%%%%%%%%%%%%%%%%%%%%%%
\section{Results}\label{sec:results} 

\begin{figure*}[t!]
    \centering
    \includegraphics[width=\textwidth]{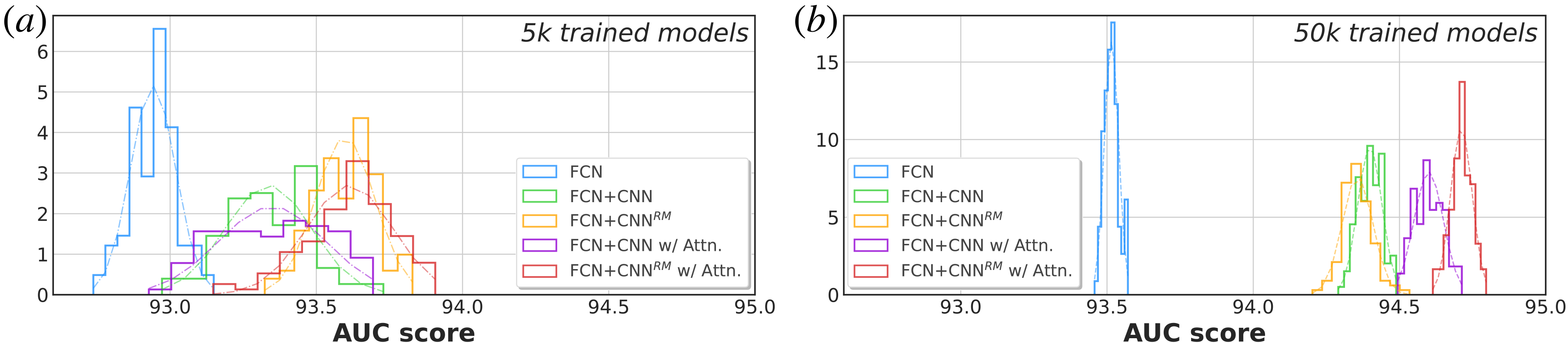}
    \caption{\label{fig:auc_set} Distribution of AUC scores for 100 independent trained models on datasets of (a) 5k and (b) 50k sizes. The histograms compare the AUC score performance across FCN (blue), FCN + CNN without (green) and with (yellow) Riemannian mapping preprocessing, and FCN + CNN with an attention layer without (purple) and with (red) Riemannian mapping preprocessing for both the training and testing phases.}
\end{figure*}
%%%%%%%%%%%%%%%%%%%%%%%%%%%%%%%%%%%%%%%%%%
With an Attention layer, we can assess the ``importance" of each contribution in details. We utilize the attention scores $a(z)$ of eq.\,(\ref{eq:att}) from the trained multi-modal model and the Attention layer across two distinct scenarios: jet images and kinematic features, utilizing a test dataset of 20 samples in FIG. \ref{fig:L2diff}. The left subplots illustrate the score distribution when L2 regularization is not applied, while the right subplots display the attention scores under the influence of L2 regularization, considering both pure jet images and Riemannian preprocessed images respectively.
%KY
The attention scores are divided with the red vertical lines along the $x$-axis into those originating from jet images (left of the red line) and those from kinematic features (right of the red line), each comprising $M=16$ weights. 
The intensity of the colors within the heatmap corresponds to the magnitude of attention scores, indicating the focus of the model on specific features. 
As previously discussed, the multi-modal model without L2 regularization shows a propensity for the attention score to favor kinematic features in the classification process. 
This bias likely arises from the data scale imbalance encountered during the extraction of local and global features by each branch.
L2 regularization is employed to mitigate this, as shown in FIG \ref{fig:L2diff}, resulting in the attention score more equitably distributing its focus between jet images and kinematic variables.

Finally, in an effort to ascertain the robustness of our models, we conducted evaluations across datasets of varying sizes, specifically those containing 5,000 (5k) and 50,000 (50k) samples, respectively.  
We consider AUC (Area Under the Curve) of ROC curves to gauge model performance in evaluating the power of a classification. 
The results of these evaluations are presented in FIG. \ref{fig:auc_set}, which illustrates the distribution of AUC scores for a suite of models: FCN, FCN combined with CNN without and with Riemannian mapping preprocessing, and their respective implications with an attention layer. 
The models equipped with attention mechanisms demonstrate an appreciable enhancement.
Moreover, the consistency in AUC score distribution across the two dataset sizes for these models highlights their stability and generalizability, a desirable attribute for applications where data availability can fluctuate significantly.
It is important to note that the underlying structures of the models employed in our analysis remain consistent throughout. 
Additionally, it is evident that when dealing with smaller training datasets such as 5k, the introduction of suitable preprocessing techniques (Riemannian mapping in our case) is crucial to ensuring the stability of the learning models. This is showcased by the improved AUC scores, indicating that the models are more reliable and provide more consistent predictions.
The hyperparameters, which are adapted in each model, are comprehensively detailed in Appendix \ref{sec:appendix2}. 
This uniformity in model structure ensures that any observed variations in performance or outcomes can be attributed primarily to the adjustments in the hyperparameters, rather than differences in the architectural frameworks of the models.

The results highlight that with an ample amount of training data, it is not only the preprocessing that contributes to model performance but also the effective integration of local and global features. The combination of these features allows the model to harness comprehensive information from both the detailed local interactions and the broader global event characteristics, enhancing the model's ability to generalize and perform robustly across varied datasets.

%%%%%%%%%%%%%%%%%%%%%%%%%%%%%%%%%%%%%%%%%%%%%%%%%%%%%%%%%%%%%%%%%%%%%%%%%%%%%%%%%%%%%%
\section{Conclusion}\label{sec:conclusion}
In this work, we introduce a novel multi-modal deep neural network architecture capable of synergistically incorporating both local (QCD) and global (kinematics) features relevant to particle physics analyses. 
By integrating kinematic features as a source of global contextual information, the model gains a comprehensive understanding crucial for classification tasks. Simultaneously, we leverage the QCD color structure to provide local, complemented attributes that serve as complementary to the global features.

Our results demonstrate that optimizing the combination of the two distinct features results in improved model prediction performance. 
This improvement shows an effective classification capability even when confronted with the inherent complexities of datasets typically encountered in particle physics experiments. 
Our model efficiently classifies signal/background events and adapts to different data scales and distributions, affirming its stability and versatility. Including attention mechanisms further refines the performance, particularly by emphasizing salient features and facilitating the identification of subtle yet essential correlations within the data. This suggests a promising direction for future research in applying deep learning techniques in high-energy physics, potentially advancing the analytical methodologies substantially.

%%%%%%%%%%%%%%%%%%%%%%%%%%%%%%%%%%%%%%%%%%%%%%%%%%%%%%%%%%%%%%%%%%%%%%%%%%%%%%%%%%%%%%%%%
\section*{Acknowledgments}
 
This work is supported by the National Research Foundation
of Korea NRF-2021R1A4A20, NRF-2019R1A2C1089334 (KB, SCP) and NRF-2021R1A2C4002551 (MP). 
KK is supported by the US DOE under Award No DE-SC0024673.

%%%%%%%%%%%%%%%%%%%%%%%%%%%%%%%%%%%%%%%%%%%%%%%%%%%%%%%%%%%%%%%%%%%%%%%%%%%%%%%%%%%%%%
%\newpage
%\newpage

\appendix

\begin{figure}[b]
    \centering
    \includegraphics[width=0.47\textwidth]{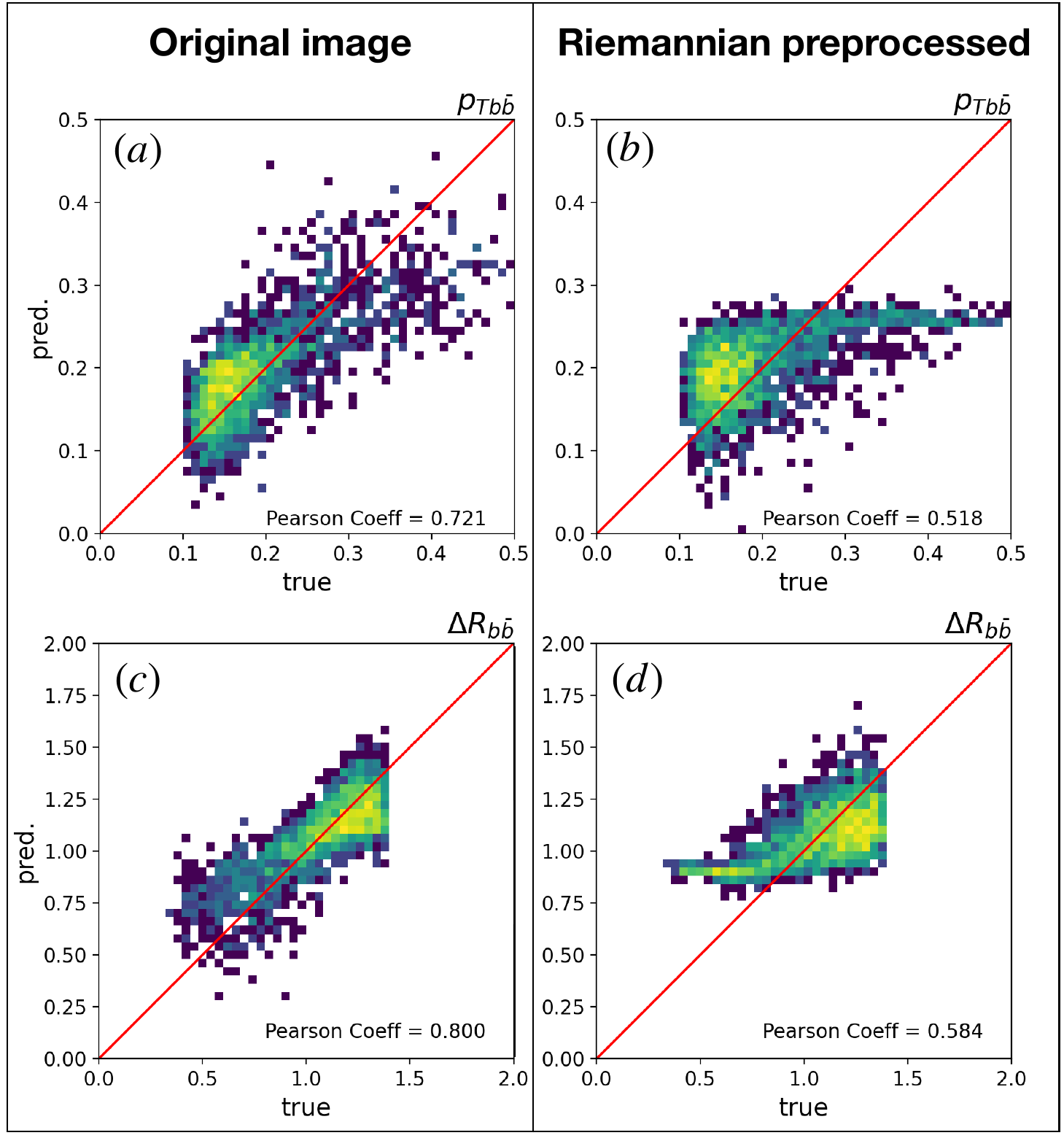}
    \caption{\label{fig:linear_regression} Linear regression analysis between predicted and true value for (a, b) $p_{Tb\bar b}$ and (c, d) $\Delta R_{b\bar b}$. Each subplot corresponds to an (a, c) original jet image and (b, d) Riemannian preprocessed one, with the $x$-axis representing the true values and the $y$-axis depicting the predicted values from the model. The red line indicates the identity line where predicted values would exactly match the true values. Pearson correlation coefficients are displayed on each plot to provide a measure of linear correlation between the true and predicted values.}
\end{figure}

\section{Decorrelating kinematic features and jet image} \label{sec:appendix1}

We describe the methodology employed to discern the degree of correlation between the kinematic features (global) and the jet images (local) of particle interactions in detail. Our procedure begins with a dimensionality reduction of the jet images, followed by linear regression to project these images into the kinematic feature space, thereby allowing us to infer the presence or absence of correlation between these two distinct feature sets.
To investigate the correlation between the kinematic features (global) and the images of jet particles (local), we followed the outlined procedure. The comparative results before and after this analysis are presented in FIG. \ref{fig:linear_regression}.

(1) For dimension reduction of the jet image ($X_{img}$), which has the shape 32×32×2, we applied Principal Component Analysis (PCA), compressing it into a 10-dimensional representation ($X_{img;PCA}$).

(2) Using Linear Regression, we regressed the PCA-reduced jet image onto the kinematic feature space as:
\[
X_{kin;train} = \mathbf{W} \times X_{img;train;PCA}
\]
If the regression of $X_{img;PCA}$ onto $X_{kin}$ is relatively unsuccessful, it can be indirectly inferred that there is a lack of correlation between the two feature sets.

(3) After getting $\mathbf{W}^*$ which is optimized results after linear regression, by using test data, we can get the 
\[
X_{kin;predicted} = \mathbf{W}^* \times X_{img;test;PCA}.
\]

(4) We then compared the distribution of the predicted kinematic features, denoted as $X_{kin;predicted}$, with the actual $X_{kin}$.

\section{Details in a neural network model}

\subsection{Preparing jet images}\label{sec:appendixjet}

\begin{figure}[b]
 \includegraphics[width=0.23\textwidth]{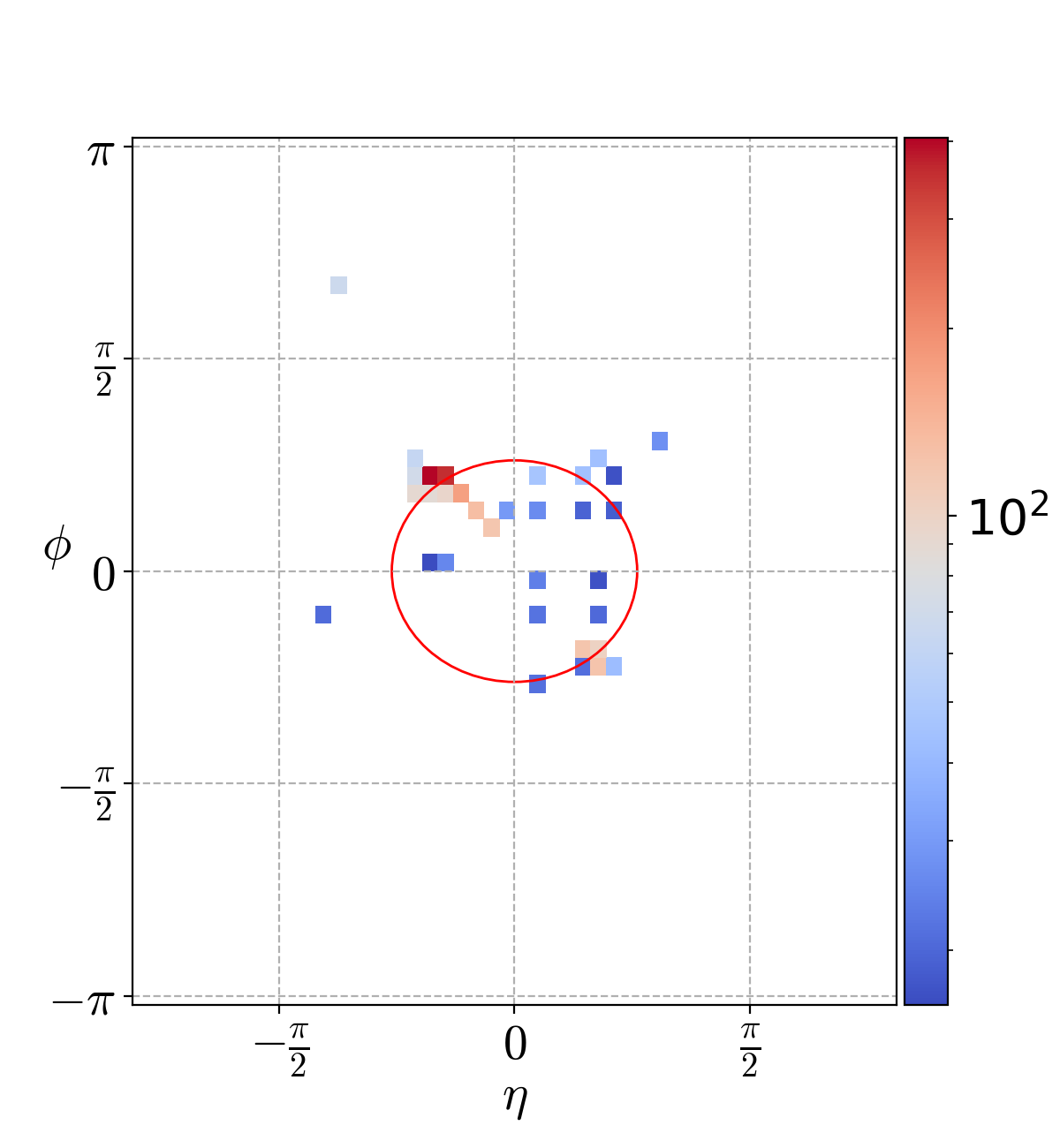} 
 \includegraphics[width=0.23\textwidth]{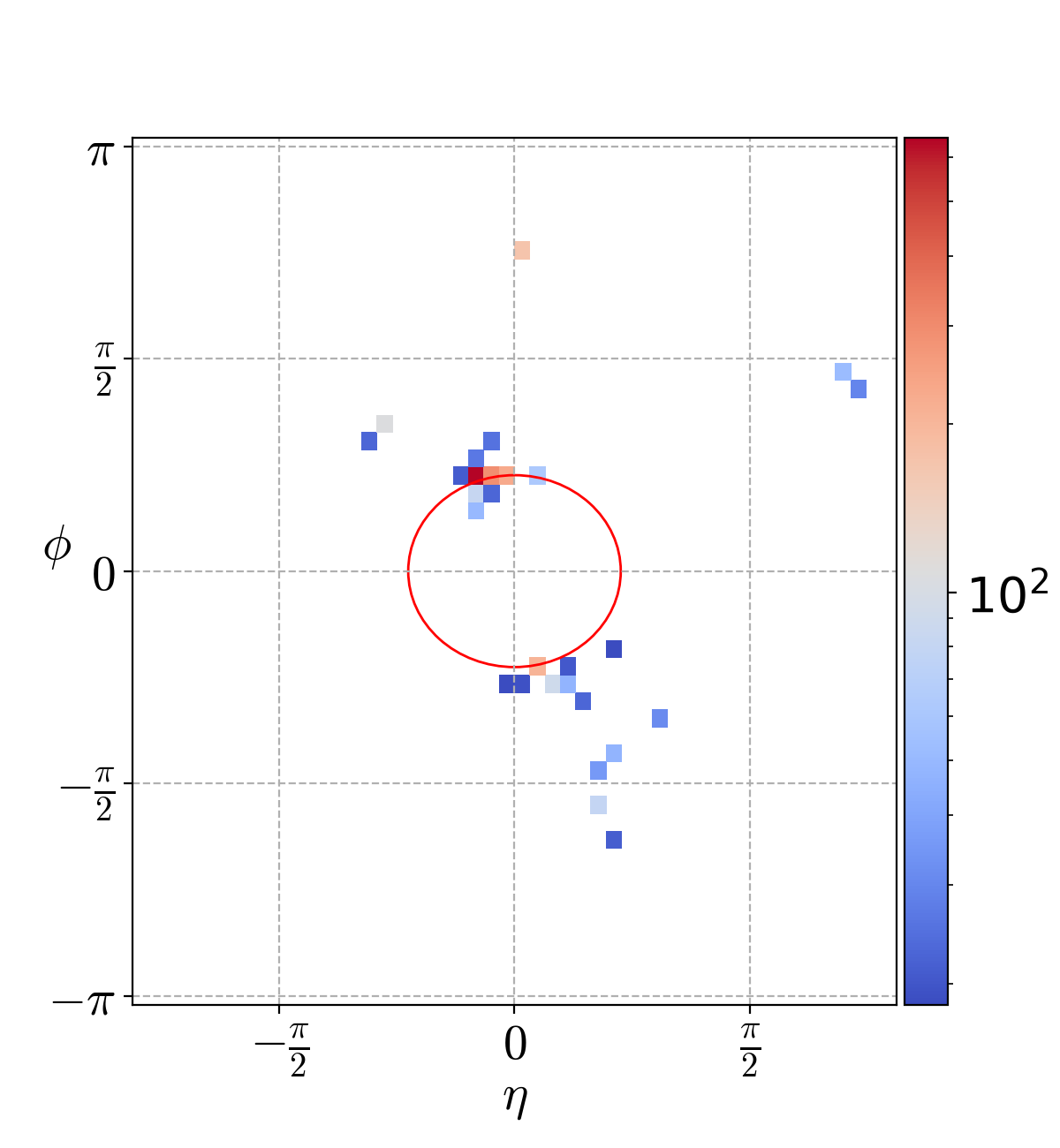} \\
 \includegraphics[width=0.23\textwidth]{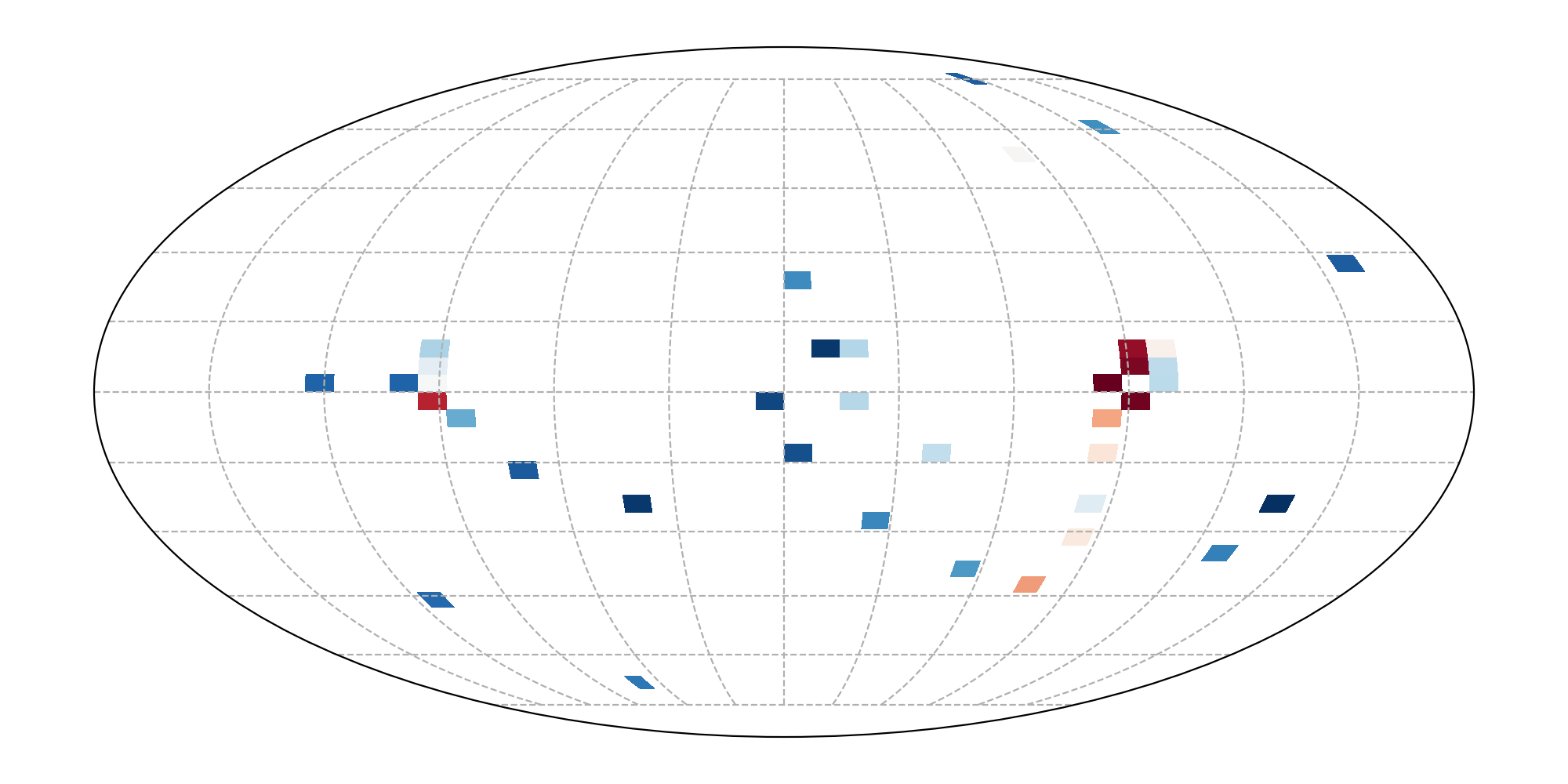} 
 \includegraphics[width=0.23\textwidth]{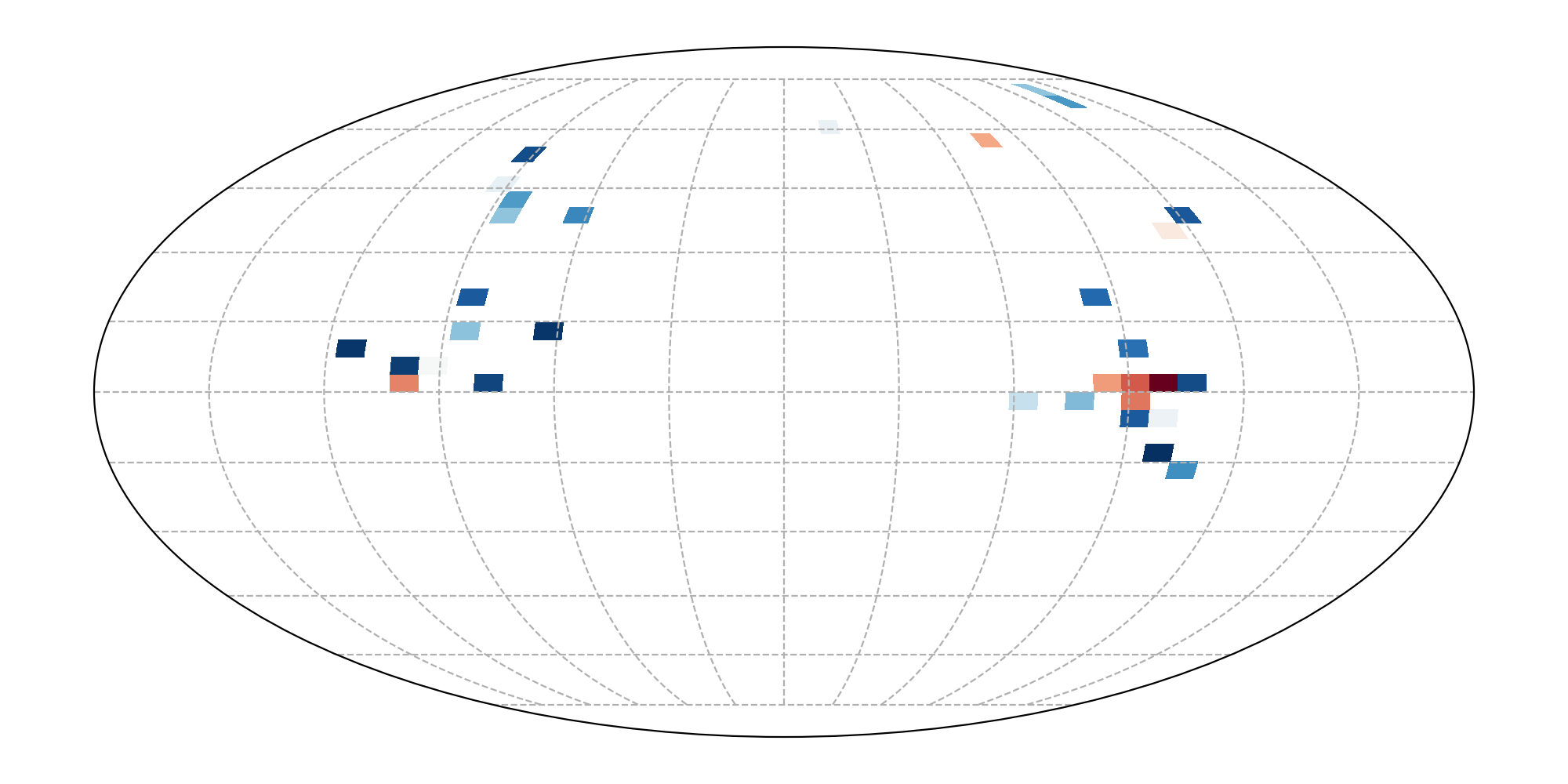}
 \caption{The transverse momentum distribution  (upper) with accumulated charged and neutral particles for a single event with $hh$ (left) and $t\bar{t}$ (right). The red circles indicate the $\Delta R_{b\bar{b}}$ of the event. The transverse momentum distribution after Riemannian preprocessing for the same event for mollweide plane (lower).}
 \label{fig:jet_img} 
\end{figure}

To prepare images for CNN, we take the conventional process which centralizes $b\bar b$-system.
FIG.\,\ref{fig:jet_img} shows the preprocessed images. 
 
\begin{enumerate}
\item {\bf{Image cleansing:}} remove all leptons and photons from the image.
\item {\bf{Particle classification:}} divide the channels into two: charged particles including charged hadrons and neutral particles including photons and neutral hadrons.
\item {\bf{Centering:}} shift the center of the image from $(0,0)$ to $(\frac{\eta_{b}+\eta_{\bar{b}}}{2},\frac{\phi_{b}+\phi_{\bar{b}}}{2})$ which is to the center of the reconstructed $b$ quark pair.
\item {\bf{Selection:}} select regions based on the $ b $ jet where $ \Delta R_{bi}>0.1 $ due to concerns that intense radiation might overshadow crucial information from soft radiation through normalization.
\item {\bf{Pixelization:}} discretize the QCD observables which are defined in ($\eta,~\phi$) plane to 50 $\times$ 50 pixels for images. We downsample the resolution of the Riemannian mapping image from 50 $\times$ 50 to 32 $\times$ 32 pixels to align with the training data size, anticipating subsequent cropping for the pure jet image.
\item {\bf{Normalization:}} scale the pixel intensity ($I_{ij}$) by $I_{ij}\rightarrow I_{ij}/\Sigma I_{ij}$, where $i$ and $j$ are the pixels index that the sum of $p_T$ in single image to be 1. 
\item  {\bf{Cropping:}} For the jet image, we crop the image with (32, 32) pixels which is related to $-1.6 < \eta <  1.6$ and $-2.01 < \phi < 2.01$ for the original jet image.
\end{enumerate}

The CNN branch consists of three convolutional layers, each followed by batch normalization, max-pooling, and a dropout layer with a rate of 0.3 to mitigate overfitting. 
Specifically, each convolutional layer employs 32 filters and a 5×5 kernel, with ReLU activation. 
Max-pooling is performed with a 2×2 window and a stride of 2.  
The output ($z_{1:M}$) is flattened and passed through a dense layer with $M$ units without activation function.

\subsection{Details in training a neural network model} \label{sec:appendix2}

\begin{table}[t!]
\setlength{\tabcolsep}{5pt}
\renewcommand{\arraystretch}{1.1}
\begin{tabular}{c|cc|cc}
\hline\hline
       Num. of training data   & \multicolumn{2}{c|}{5k}        &  \multicolumn{2}{c|}{50k}   \\ \hline
                        & \multicolumn{1}{c|}{$\ell$} &  $l_2$  &  \multicolumn{1}{c|}{$\ell$} & $l_2$ \\ \hline\hline
FCN 10 vars             & \multicolumn{1}{c|}{$5\times10^{-3}$} & -     & \multicolumn{1}{c|}{$5\times10^{-3}$}               & -     \\ \hline
FCN+CNN                 & \multicolumn{1}{c|}{$5\times10^{-3}$} & -     & \multicolumn{1}{c|}{$10^{-2}$}             & -     \\ \hline
FCN+CNN$^{RM}$          & \multicolumn{1}{c|}{$\times10^{-2}$}        & -     & \multicolumn{1}{c|}{$10^{-2}$}            & -     \\ \hline
FCN+CNN w/ Attn.        & \multicolumn{1}{c|}{$5\times10^{-3}$}        &  $10^{-2}$     &  \multicolumn{1}{c|}{$10^{-2}$}            &   $10^{-2}$    \\ \hline
FCN+CNN$^{RM}$ w/ Attn. & \multicolumn{1}{c|}{$10^{-2}$} &    $10^{-2}$   &  \multicolumn{1}{c|}{$5\times10^{-3}$}           &     $10^{-2}$  \\ \hline\hline
\end{tabular}
\caption{Selected hyperparameters for each model, as referenced in the results presented in FIG. \ref{fig:auc_set}. The table delineates the optimized values for the learning rate $( \ell $) and L2 regularization coefficient ($ l_2$) corresponding to the training datasets comprising 5,000 (5k) and 50,000 (50k) samples.}
\label{tab:hyper} 
\end{table}

The classification part after an attention block consists of three dense layers consists with 64 units, as indicated by the parameter $M$ set to 16.
These layers employ the ReLU activation function, and between these dense layers, batch normalization is applied to stabilize and accelerate the learning process by normalizing the activations. 
Finally, the network culminates in an output layer with a single unit employing a sigmoid activation function, which is standard for binary classification tasks. 

 For the DNN library, we use \textsf{Keras} package \cite{chollet2015} and the models are constructed with \textsf{ReLU} activations and we use \textsf{Adam} optimizer \cite{Adam} to minimize the binary cross-entropy loss function. To ensure stable training, we employ a learning rate scheduler that maintains a constant learning rate for the initial 10 epochs. Subsequently, the learning rate is adjusted according to the formula $\ell\times e^{-\beta}$, where $\beta$ represents the decay factor, and we set $\beta = 0.5$.
 
We investigate the robustness of our model with respect to varying training dataset sizes to highlight its stability. Specifically, we conduct evaluations across datasets of 5,000 and 50,000 samples. In this process, we perform a thorough optimization of parameters for each dataset and corresponding model configuration. The optimal set of hyperparameters is determined based on achieving the highest average Area Under the Curve (AUC) scores over 10 independent runs. We select the adjusted hyperparameters from a predefined range, with learning rates $\ell$ chosen from \{0.01, 0.005, 0.001\} and L2 regularization strengths  $l_2$ from \{0.5, 0.1, 0.05, 0.01\}. The hyperparameters that result from this optimization are documented in TABLE \ref{tab:hyper}.

\bibliography{ref}

\end{document}